\documentclass[
 reprint,
superscriptaddress,
 amsmath,amssymb,
 aps,
]{revtex4-2}

\usepackage{graphicx}
\usepackage{dcolumn}
\usepackage{bm}
\usepackage{hyperref}
\usepackage{xcolor}

\newcommand{\mean}[1]{\langle #1 \rangle}

\newcommand{\Eqref}{Eq.~\eqref}
\newcommand{\Figref}{Fig.~\ref}
\newcommand{\nodes}{N_\mathrm{nodes}}

\begin{document}

\title{Memory capacity of adaptive flow networks}

\author{Komal Bhattacharyya}
\affiliation{%
Max Planck Institute for Dynamics and Self-Organisation, 37077 Göttingen, Germany
}%
 \author{David Zwicker}
 \affiliation{%
Max Planck Institute for Dynamics and Self-Organisation, 37077 Göttingen, Germany
}%
 \author{Karen Alim}%
 \email{k.alim@tum.de}
\affiliation{%
Max Planck Institute for Dynamics and Self-Organisation, 37077 Göttingen, Germany
}%
\affiliation{Physik-Department and Center for Protein Assemblies, Technische Universität München, 85748 Garching, Germany}

\date{\today}

\begin{abstract}
Biological flow networks adapt their network morphology to optimise flow while being exposed to external stimuli from different spatial locations in their environment. These adaptive flow networks retain a memory of the stimulus location in the network morphology. Yet, what limits this memory and how many stimuli can be stored is unknown. Here, we study a numerical model of adaptive flow networks by applying multiple stimuli subsequently. We find strong memory signals for stimuli imprinted for a long time into young networks. Consequently, networks can store many stimuli for intermediate stimulus duration, which balance imprinting and ageing. 
\end{abstract}

\maketitle

\section{\label{sec:level1}Introduction}
Biological flow networks, like vasculature~\cite{Murray1926}, fungal mycelium~\cite{Boddy2010}, or slime mold~\cite{Marbach2016,Alim2013}, optimise their function  
by remodelling the network morphology in response to internal and external stimuli~\cite{Sugden2017,Chen2012,Hu2012,Tero2008,Tero2010,Marbach2016,Marbach2022}. In particular, the slime mold \textit{Physarum polycephalum} reorganises its network morphology during foraging and migration~\cite{Kamiya1988,Kuroda2015}, or as responses to environmental influences~\cite{Kramar2021a}, by adapting tubes to flow~\cite{Tero2008,Tero2010,Alim2017b}. Although the organism only consist of a single cell, it processes information~\cite{Gao2019,Beekman2015,Tero2010,Tero2006,Tero2008,Nakagaki2000} and stores memory of external stimuli in the network morphology~\cite{Kramar2021a,Bhattacharyya2022}. 
Yet, the information processing capabilities of \textit{Physarum} in particular, and adaptive flow networks more generally, are so far unclear.

The self-organised information processing of \textit{Physarum} is reminiscent of other physical learning systems~\cite{Stern2022}:
Physical networks can be trained to have unusual mechanical properties \cite{Reid2018,Pashine2019} and functionalities~\cite{Rocks2017,Hexner2020}
 either by modifying microscopic properties by global optimisation~\cite{Yan2017} or 
as local responses~\cite{Dillavou2021}.
Such networks can also learn multiple states \cite{Rocks2019}, which is key for obtaining multi-functionality~\cite{Rocks2019} and multi-stability~\cite{Yang2018,Steinbach2016,Che2017,Bertoldi2017,Overvelde2016,Silverberg2015,Waitukaitis2015,Yang2018a,Fu2018,Kim2019} in physical systems, and for performing complex tasks like image classification~\cite{Stern2020,Stern2021,Anisetti2022} using such physical networks.
The multiple states can be either imprinted simultaneously~\cite{Rocks2019} or learned subsequently~\cite{Stern2020}. In both cases, there is a maximal number of states that can be learned, which is the learning capacity of the system~\cite{Rocks2019,Stern2020}.

Although memory is essential for learning~\cite{Stern2022}, the memory capacity of flow networks remains unexplored.
We, here, investigate this question theoretically, by analysing memory in a model of adaptive flow network, which is subjected to various external stimuli, similar to natural flow networks~\cite{Ito2010,Meyer2017}.
We identify that a stimulus is stored more robustly, and can, thus, be retrieved more easily, when networks are young and are exposed to a stimulus for a long time.
Since these two criteria are incommensurable for multiple stimuli, a trade-off determines the memory capacity of these adaptive flow networks.

\section{\label{level2}Model}
\begin{figure*}[t!]
\centering
\includegraphics[width=\textwidth]{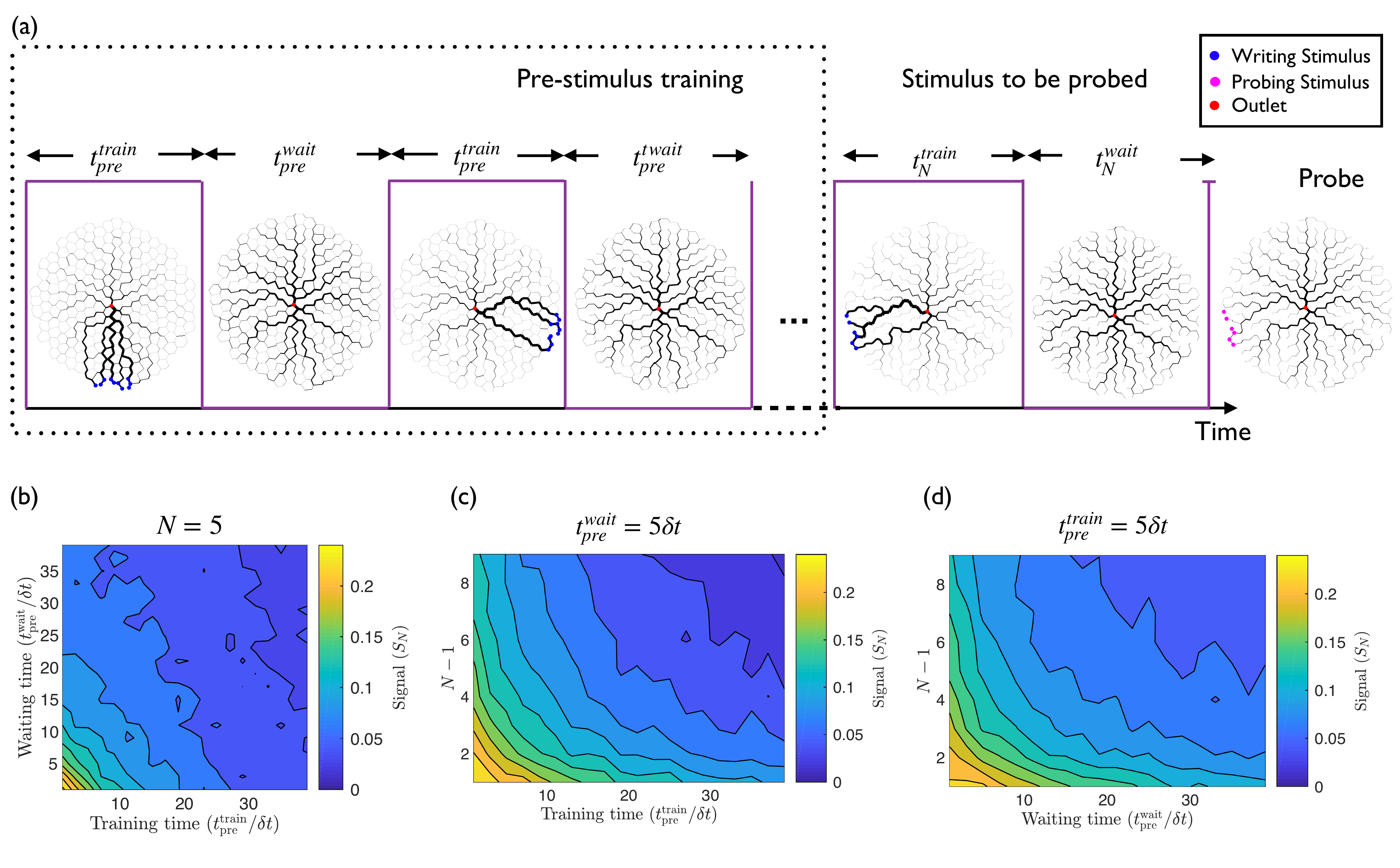}
\caption{Stimulus locations are retained in previously stimulated networks.
(a)~Schematic of adaptive flow networks with a central outlet (red dot), fluctuating inflows at all other nodes, and additional inflow for stimuli (blue dots).
The temporal sequence shows snapshots of the training protocol, where $N$ stimuli (purple boxes) are applied sequentially by stimulating for a period $t^\mathrm{train}_n$ followed by relaxation period~$t^\mathrm{wait}_n$ for each stimulus.
Finally, the last stimulus is probed (pink dots) to determine the signal~$S_N$ according to \Eqref{eq:signal}.
(b)~$S_N$ as a function of the training time $t^\mathrm{train}_n = t^\mathrm{train}_\mathrm{pre}$ and waiting time $t^\mathrm{wait}_n=t^\mathrm{wait}_\mathrm{pre}$ for most stimuli ($n<N$) with $N=5$.
(c)~$S_N$ as a function of $t^\mathrm{train}_\mathrm{pre}$ and $N$ for $t^\mathrm{wait}_\mathrm{pre}=5\delta t$.
(d)~$S_N$ as a function of $t^\mathrm{wait}_\mathrm{pre}$ and $N$ for $t^\mathrm{train}_\mathrm{pre}=5\delta t$.
(a--d)
Model parameters are $t^\mathrm{train}_N=10\delta t$, $t^\mathrm{wait}_N=5\delta t$, $\nodes=1100$, $q^\mathrm{add} = 2000\,q^{(0)}$, $q^{(0)} = 1$, $\mathcal K=1600$, and $T = 30\delta t$. Data shows mean from $1500$ independent simulations.}
\label{fig:1}
\end{figure*}
We use the standard model for adaptive flow networks that minimises energy dissipation in order to maximise transport through the network, for fixed network building material~\cite{Murray1926,Hacking1996,Bohn2007,Corson2010,Katifori2010,Hu2013}.
These flow networks are modelled as a graph of $\nodes$ nodes connected by links ${ij}$, where $i,j\in \{1,\ldots,\nodes\}$.
The links have length $l_{ij}$ and time-dependent conductances $C_{ij}(t)$.
We consider a network of cylindrical hollow tubes with conductances $C_{ij} = \pi r_{ij}(t)^4/8\mu l_{ij}$ according to Hagen-Poiseuille's law, where $\mu$ is the viscosity of the enclosed fluid.
In our case, node $i=1$ serves as the sole outlet, while all other nodes are inlets with fluctuating inflows $q_i(t)$, which are either $q_i=0$ or $q_i=2q^{(0)}$ with equal probability.
Conservation of total flow implies $q_1(t) = -\sum_{i>1} q_{i}(t)$.
We chose a disk-shaped network geometry with the outlet in the centre; see Fig.~\ref{fig:1}(a).
Conservation of flow at every node, as described by Kirchhoff's law, then uniquely determines the flow $Q_{ij}(t)$ in all links, given the entire networks conductances $C_{ij}(t)$ and the inflows $q_i(t)$; see Appendix A.
The adaptive dynamics follow from the assumption that networks minimise dissipation~\cite{Murray1926}
\begin{equation}
    E(t)=\sum_{<ij>}\frac{Q_{ij}(t)^{2}}{C_{ij}(t)}
    \;,
\end{equation} 
while obeying the constraint
\begin{equation}
  \label{eqn:constraint}
   \mathcal{K}^{\frac12} =  \sum_{<ij>} C_{ij}(t)^{\frac12}l_{ij}^{\frac32}
   \;,
\end{equation}
where $\mathcal{K}^{\frac12}$ is proportional to the fixed overall volume of all links.
We follow an iterative relaxation algorithm~\cite{Bohn2007}, where the conductances at the next time step, $C_{ij}(t+\delta t)$, 
adapt to minimise $E(t)$ while obeying \Eqref{eqn:constraint}, implying
\begin{equation}\label{eq:iteration}
    C_{ij}(t+\delta t) = \frac{\mathcal{K}\mean{Q_{ij}(t)^2}_T ^{\frac{2}{3}}}
    	{\left(\sum _{<ij>}\langle Q_{ij}(t)^{2}\rangle_T^{\frac{1}{3}}l_{ij}\right)^{2}l_{ij}}
	\;,
\end{equation}
where we average the flow over a duration $T$, $\mean{Q_{ij}^2}_T$, since the inflows at every node fluctuate over time.

To probe for memory, we initiate networks with conductances~$C_{ij}$ chosen uniformly from the interval $[0, 1]$, which are then rescaled, so they obey the constraint given by \Eqref{eqn:constraint}.
We then stimulate the networks using an additional inflow $q^{\mathrm{add}}$ at the outer rim at a specific angular location; see Fig.~\ref{fig:1}(a).
We distribute the additional inflow over a few nodes to avoid artefacts from the symmetries of the underlying networks.
The adaptation dynamics then imprint the stimulus in a treelike structure from the nodes of additional inflow to the centred outlet \cite{Bhattacharyya2022}; see Fig.~\ref{fig:1}(a).
Once the additional inflow is withdrawn, networks return to seemingly isotropic morphologies.
Yet, when probing networks by re-applying an additional load at exactly the same location, the power loss of previously stimulated, and thus trained, networks, $E_{\mathrm{trained}}$, is distinctively less than if probed at any other location.
In particular, $E_{\mathrm{trained}}$ is less than the power loss $E_{\mathrm{untrained}}$ for probing untrained networks that evolved for the same total time, but did not see the stimulus~\cite{Bhattacharyya2022}.
To quantify this memory, we established the normalised difference in power loss between trained and untrained networks as a measure of the memory readout signal $S$, \cite{Bhattacharyya2022},
\begin{equation}
\label{eq:signal}
    S = 1 - \frac{\mean{E_{\mathrm{trained}}}}{\mean{E_{\mathrm{untrained}}}}
    \;,
\end{equation}
where brackets indicate ensemble averages over initial configurations and positions of the additional loads.
We used this quantification to show that freshly initiated networks memorise single stimuli in the spatial location and orientation of the vanishing links~\cite{Bhattacharyya2022}.
The stimulus can be read out by probing the network again
after the stimulus is withdrawn.
Yet, it is unclear how well already stimulated networks can store stimuli and whether networks can store multiple stimuli simulatenously.

\section{\label{sec:level3}Results}
\subsection{\label{sec:level3a} Pre-stimulated networks can memorise stimuli}
We start by asking whether previously evolved networks can store a stimulus reliably.
We evolve networks with a stimulation protocol by consecutively applying $N$ stimuli, distinguished by the angle of the additional inflow.
We choose the angles randomly from $10$ possibilities, $\{0, \frac{\pi}{5}, \ldots, \frac{9\pi}{5}\}$, and we set the angular range of each stimulus to $\frac{\pi}{6}$ to avoid stimuli overlap.
Starting with a randomly initialised network, we apply one stimulus after the other.
The $n$-th stimulus is imprinted on the network by iterating \Eqref{eq:iteration} with the additional load corresponding to the stimulus for a duration $t^\mathrm{train}_n$ and then without load for a duration~$t^\mathrm{wait}_n$.
Taken together, the network evolved to time
\begin{equation}
\label{eq:age}
    t^\mathrm{age}_n = \sum_{m=1}^n \left(t^\mathrm{train}_m + t^\mathrm{wait}_m\right)
\end{equation}
after the $n$-th stimulus has been applied.

To test whether a stimulated network can memorise an additional stimulus, we apply $N-1$ stimuli with identical properties and then probe the signal of a final stimulus; see Fig.~\ref{fig:1}(a).
We thus have $t^\mathrm{train}_n = t^\mathrm{train}_\mathrm{pre}$ and $t^\mathrm{wait}_n = t^\mathrm{wait}_\mathrm{pre}$ for $n<N$, while the final stimulus can have different parameters.
The signal~$S_N$ quantifies the dissipation difference of applying the $N$-th stimulus, analogously to \Eqref{eq:signal}.
For constant parameters of the pre-stimulation protocol, we observe that $S_N$ increases with $t^\mathrm{train}_N$ and decays with $t^\mathrm{wait}_N$; see Appendix B.
This behaviour closely resembles memory formation in a freshly initiated network~\cite{Bhattacharyya2022}, even though we here use pre-stimulated networks.

We next test the influence of the precise pre-stimulation protocol by varying the number of applied stimuli, $N$, the training time, $t^\mathrm{train}_\mathrm{pre}$, and the relaxation time, $t^\mathrm{wait}_\mathrm{pre}$.
\Figref{fig:1}(b--d) shows that the signal~$S_N$ of the final stimulus decreases when increasing any of these parameters, so the pre-stimulation protocol affects how well additional memories can be stored.
However, our simulations demonstrated that pre-stimulated adaptive networks can store information about additional stimuli.

\subsection{\label{sec:level3c}Memory capacity reduces with age}
\begin{figure}
\centering
\includegraphics[width=0.5\textwidth]{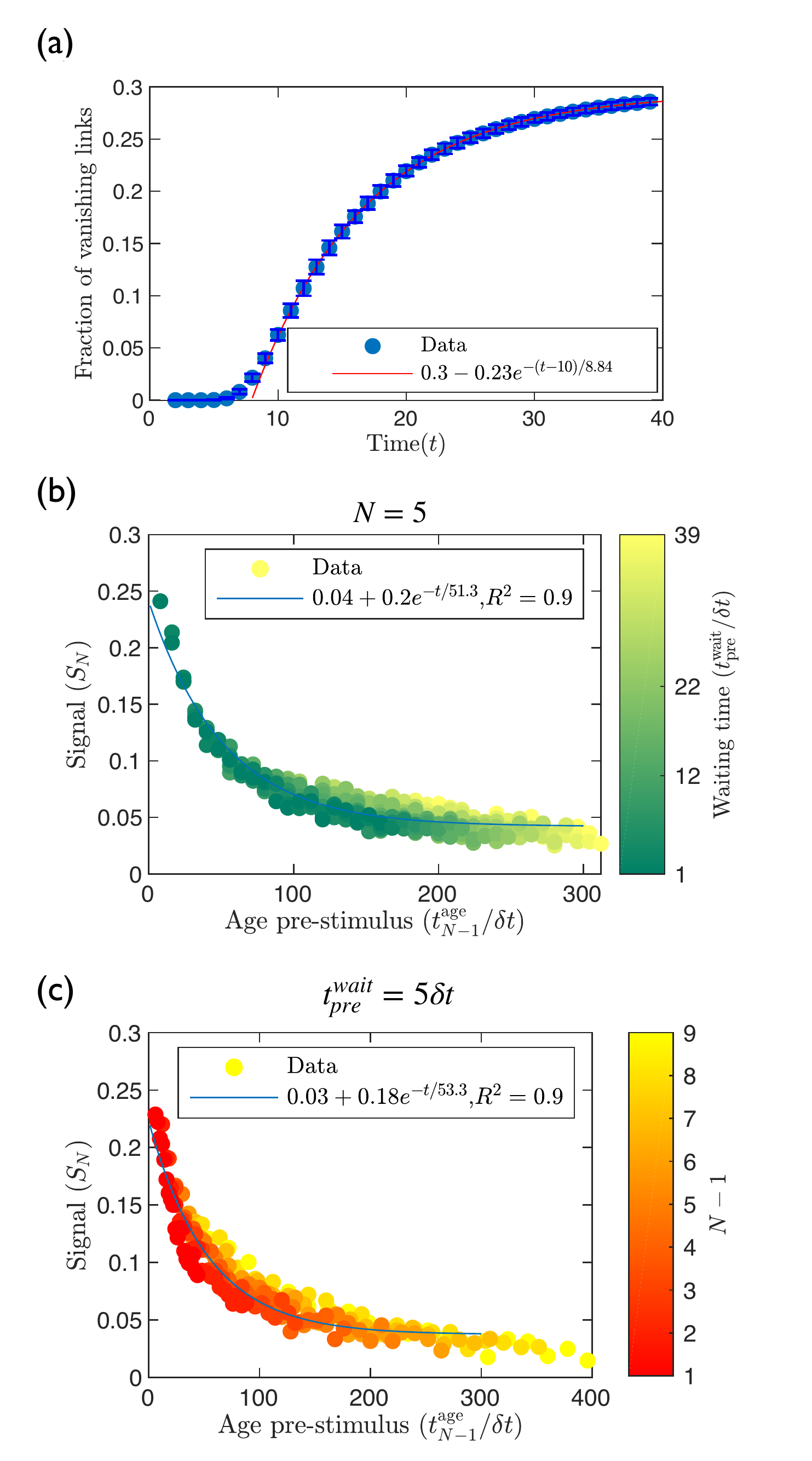}
\caption{\label{fig:2} Memory signal of final stimulus reduces with network age.
(a)~Fraction of vanishing links (blue symbols) as a function of network iterations averaged over $80$ independent runs.
Model parameters $\nodes$, $q^{\mathrm{add}}$, $q^{(0)}$, and $T$ are given in \Figref{fig:1}.
Red line indicates an exponential fit.
(b, c) Signal~$S_N$ of final stimulus as a function of age~$t^\mathrm{age}_{N-1}$ before stimulus was applied.
Panels b and c show data of \Figref{fig:1}(b) and \Figref{fig:1}(c), respectively.
Blue lines indicate exponential fits.
}
\end{figure}
\begin{figure*}
\centering
\includegraphics[width=\textwidth]{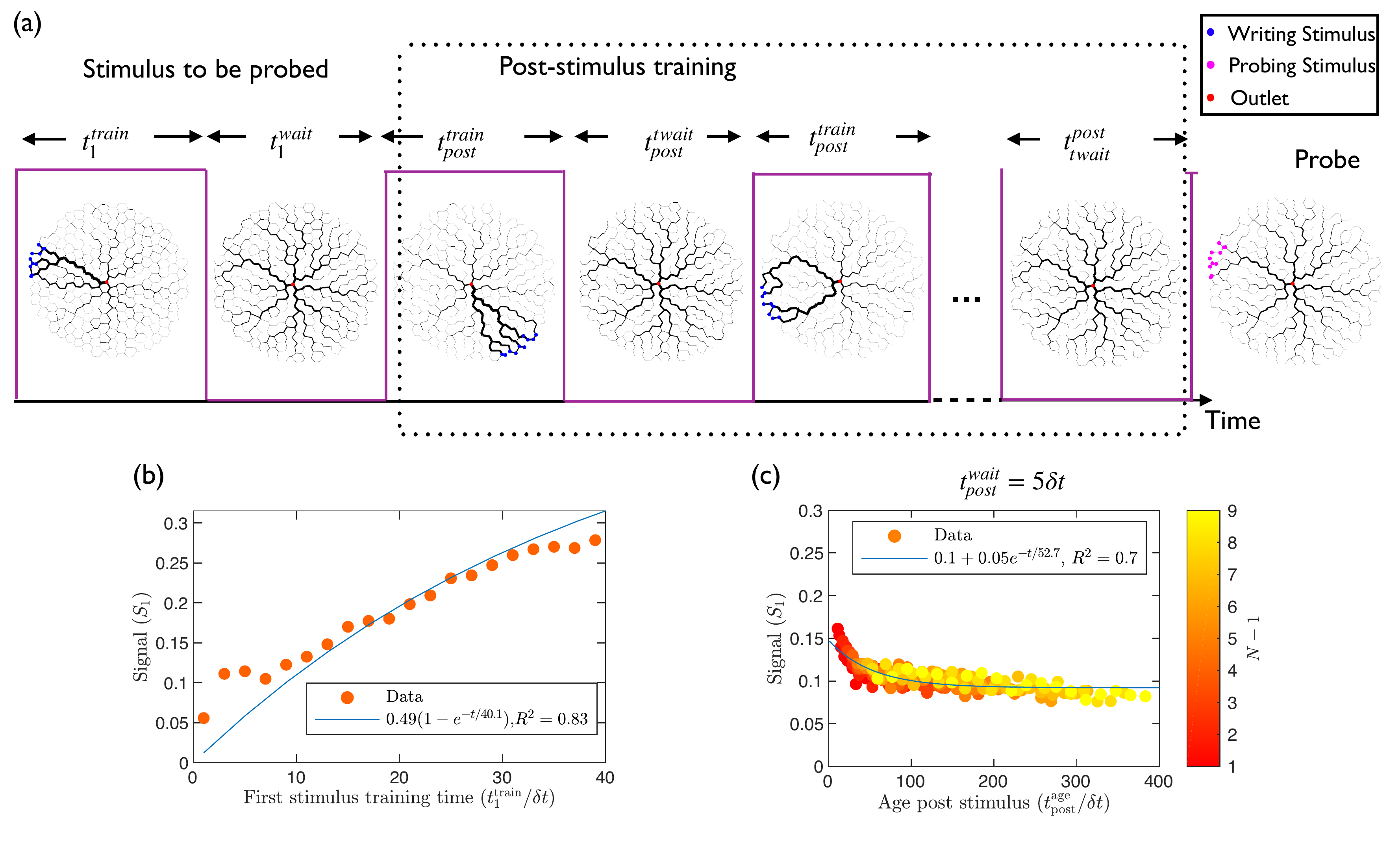}
\caption{\label{fig:3} Signal~$S_1$ of first stimulus increases with training time and decreases with network age. 
(a)~Snapshots of network, which is subjected to the first stimulus for $t^\mathrm{train}_1$, relaxed for $t^\mathrm{wait}_1$, and then $N-1$ stimuli are applied with $t^\mathrm{train}_n=t^\mathrm{train}_\mathrm{post}$ and $t^\mathrm{wait}_n = t^\mathrm{wait}_\mathrm{post}$ ($n>1$, dotted box), until the first stimulus is  probed.
(b)~$S_1$ as a function of $t^\mathrm{train}_1$ for $N=5$ and $t^\mathrm{train}_\mathrm{post}=5\delta t$.
(c)~$S_1$ as a function of $t^\mathrm{age}_\mathrm{post}=t^\mathrm{age}_N - t^\mathrm{train}_1$ for various $N$ at $t^\mathrm{train}_1 =10\delta t$.
(b--c) Blue lines indicate exponential fits. Parameters are $t^\mathrm{wait}_1 =t^\mathrm{wait}_\mathrm{post} =5\delta t$ and given in \Figref{fig:1}.}
\end{figure*}
We next investigate how the pre-stimulation protocol affects the memory of the final stimulus.
Since information about stimuli locations are stored in the orientation and location of irreversibly decaying links~\cite{Bhattacharyya2022}, we first determine how the micro-structure of the network evolves with time.
\Figref{fig:2}(a) shows that the average fraction of vanishing links saturates exponentially with time, which suggest that the memory capacity of adaptive flow networks decreases with the time~$t^\mathrm{age}_{N-1}$, given by \Eqref{eq:age}, that the network evolved for before the stimulus is applied.
Re-plotting the memory signal $S_N$ of the final stimulus as a function of $t^\mathrm{age}_{N-1}$ leads to a data collapse for various values of $N$, $t^\mathrm{train}_\mathrm{pre}$, and $t^\mathrm{wait}_\mathrm{pre}$; see \Figref{fig:2}(b, c).
The two panels differ in whether $N$ (panel b) or $t^\mathrm{wait}_\mathrm{pre}$ (panel c) are kept fixed while the other parameters are varied.
In both cases, the data collapse is well-described by an exponential decay
\begin{equation}
\label{eq:signal_final}
    S_N(t^\mathrm{age}_{N-1}) \approx S^{\infty}_N
    	+A_N\exp\left(-\frac{t^\mathrm{age}_{N-1}}{\tau_\mathrm{pre}}\right)
	\;,
\end{equation}
where $\tau_\mathrm{pre}\approx 52\,\delta t$ denotes the time scale, with which pre-stimulation reduces the memory capacity of the final stimulus.
The maximal memory capacity, $S^{\infty}_N + A_N \approx 0.22$ for $t^\mathrm{age}_{N-1}=0$, is significantly larger than the residual capacity, $S^{\infty}_N\approx 0.03$, consistent with the fact that pre-stimulation of the networks reduces the memory capacity.
The fact that the exponential decay adequately describes the decreasing capacity suggests that only the total duration of pre-stimulation is important, while the details of the protocol are irrelevant.
Consequently, younger networks allow for a larger memory signal of the final stimulus.

\subsection{\label{sec:level3d}Training time dominates signal of first stimulus}
\begin{figure*}
\centering
\includegraphics[width=\textwidth]{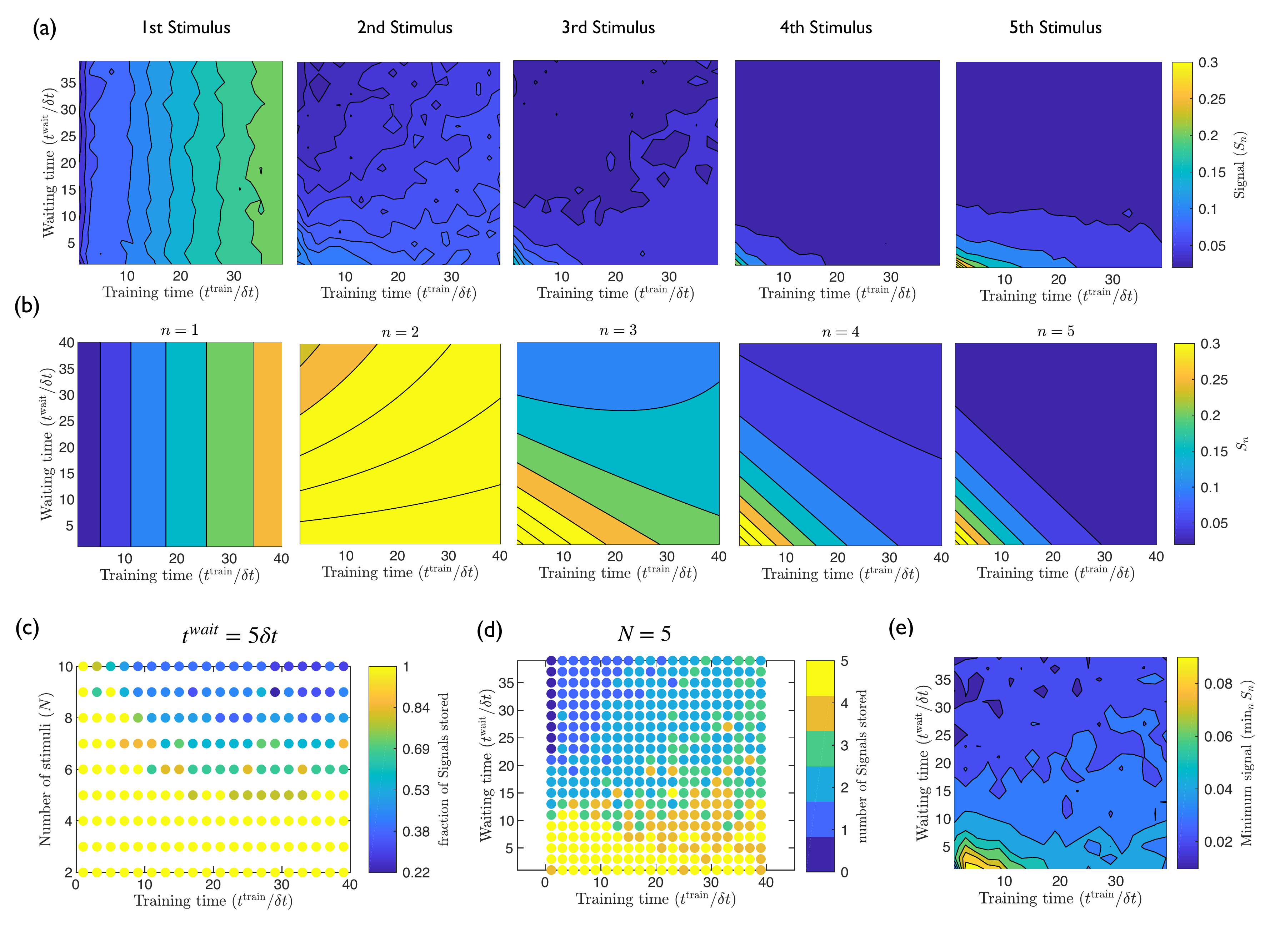}
\caption{\label{fig:4}Memory capacity depends on stimulation protocol parameters.
(a)~Numerically obtained signals~$S_n$ for all $N=5$ stimuli as functions of training times $t^\mathrm{train}$ and waiting times $t^\mathrm{wait}$.
(b)~Analytical prediction of $S_n$ given by \Eqref{eq:multsignal} for all stimuli as functions of $t^\mathrm{train}$ and $t^\mathrm{wait}$ for $N=5$.
(c)~Fraction of stimuli with strong signal ($S_n>0.04$) as a function of $t^\mathrm{train}$ and $N$ for $t^\mathrm{wait} =5\delta t$. 
(d)~Number of stimuli with $S_n>0.04$ as a function of $t^\mathrm{train}$ and $t^\mathrm{wait}$ for $N =5$. 
(e)~Minimal signal $\min_n S_n$ of $N=5$ stimuli as a function of $t^\mathrm{train}$ and $t^\mathrm{wait}$.
(a--e) Model parameters are given in \Figref{fig:1}.
 }
\end{figure*}
To retain multiple memories, adaptive networks need to store information about all stimuli.
We, thus, next investigate how information about earlier stimuli are retained and particularly focus on the first stimulus.
To investigate the first stimulus in detail, we change the protocol to control the training parameters of the first stimulus separately from all the other stimuli; see \Figref{fig:3}(a).
For simplicity, we use identical parameters for the other stimuli, $t^\mathrm{train}_n = t^\mathrm{train}_\mathrm{post}$ and $t^\mathrm{wait}_n = t^\mathrm{wait}_\mathrm{post}$ for $n=2,..,N$.
The network is probed at the same location as the first stimulus to obtain the memory signal~$S_1$ of the first stimulus. 
Fig.~\ref{fig:3}(b) shows that $S_1$ increases with the training time of the first stimulus, $t^\mathrm{train}_1$, and approaches zero for $t^\mathrm{train}_1 = 0$. 
$S_1$ again shows an exponential saturation,
\begin{equation}
\label{eq:signal_first}
    S_1(t^\mathrm{train}_1) \approx B_1 \left[1-\exp\left(-\frac{t^\mathrm{train}_1}{\tau_\mathrm{train}}\right)\right]
    \;,
\end{equation}
where $\tau_\mathrm{train}$ is the training time scale and $B_1$ denotes the maximal signal for $t^\mathrm{train}_1 \rightarrow \infty$.
Similar to our previous work~\cite{Bhattacharyya2022}, longer training leads to a stronger signal.

We next investigate how the signal of the first stimulus depends on subsequently applied stimuli.
\Figref{fig:3}(c) indicates that $S_1$ decays as the networks evolves further, similar to our previous study~\cite{Bhattacharyya2022}.
We find that $S_1$ only depends on the duration of evolution after the first training period, $t^\mathrm{age}_\mathrm{post}  = t^\mathrm{age}_N - t^\mathrm{train}_1$, and not the precise details of the protocol.
Moreover, $S_1$ again decays exponentially,  
\begin{equation}
    S_1(t^\mathrm{age}_\mathrm{post}) \approx S^{\infty}_1 + A_1\exp\left(-\frac{t^\mathrm{age}_\mathrm{post}}{\tau_\mathrm{post}}\right)
    \;,
\end{equation}
where the coefficients have the same interpretation as in \Eqref{eq:signal_final}.
Our fits indicate that $\tau_\mathrm{post} \approx \tau_\mathrm{pre}$, consistent with an intrinsic time scale of memory formation.
Note that the residual memory capacity $S^\infty_1 \approx 0.1$ is large, implying that subsequent training does not affect the signal very strongly.
This is consistent with the picture that memory is stored by vanishing links that cannot be revived; see analytical and numerical observations of the transitions phase between stimuli in Appendix C and D.
Taken together, we find that adaptive flow networks can store multiple memories.   

\subsection{\label{sec:level3e}Trade off between age and training time limits memory capacity}
We found that stimuli are imprinted most strongly when they are trained for a long time on a young network. These goals of long training times and young networks are contradictory for late stimuli, suggesting there must be a trade-off for best performance of imprinting multiple stimuli. To understand how many stimuli can be imprinted in a network, we next consider $N$ non-overlapping stimuli with identical stimulation parameters, $t^\mathrm{train}_n = t^\mathrm{train}$ and $t^\mathrm{wait}_n = t^\mathrm{wait}$ for $n=1,..,N$.  
We now also probe all stimuli locations to obtain a signal $S_n$ for each stimulus.
\Figref{fig:4}(a) shows data for five stimuli as a function of $t^\mathrm{train}$ and $t^\mathrm{wait}$.
We recover that the signal~$S_1$ of the first stimulus mainly depends on the training time~$t^\mathrm{train}$ and is barely affected by the subsequent dynamics.
Conversely, the signal of all other stimuli decreases with network age, i.e., with increasing $t^\mathrm{train}$ and $t^\mathrm{wait}$.
In particular, mid-timed stimuli have the weakest signals, suggesting that they are affected by both pre-stimuli ageing as well as subsequent degradation.

We next develop an analytical prediction of the signal of all stimuli, motivated by the successful description of the signals of the first and last stimulus demonstrated above.
We hypothesise that the signal of the $n$-th stimulus is a combination of the pre-stimulus ageing, described by \Eqref{eq:signal_final}, and the actual training, described by \Eqref{eq:signal_first}, while we neglect the small effect of the post-stimulus signal degradation.
We show in Section V of the Supplementary Information that a weighted sum of the two effects adequately describes the data, which results in the prediction
\begin{equation}
\label{eq:multsignal}
    S_n \approx \frac{1}{2} \begin{cases}
    1 - e^{-\frac{t^\mathrm{train}}{\tau_\mathrm{train}}}  & n = 1
\\
    	e^{-\frac{t^\mathrm{age}_{n-1}}{\tau_\mathrm{pre}}} 
	+ \left(\frac{n}{2}\right)^{1-n}\left(1-e^{-\frac{t_\mathrm{train}}{\tau_\mathrm{train}}}\right)
	&  n>  1 \;,
 \end{cases}
\end{equation} 
where $t^\mathrm{age}_n$ is given by \Eqref{eq:age}.
This equation correctly captures that $S_n$ decreases with the $n-1$ previously applied stimuli.
\Figref{fig:4}(b) shows that \Eqref{eq:multsignal} also captures the qualitative features of the dependence on $t^\mathrm{train}$ and $t^\mathrm{wait}$.
However, the prediction overestimates the signal of mid-timed stimuli, likely because we neglect the post-stimuli degradation. We also note that this analytical description does not reproduce the quantitative features of the signal observed numerically because the signal is not just a linear superposition of the training and age impact; see Appendix E. Even though more research is needed to obtain the exact dependency of the signal on training and age, we choose to use the simple function to draw insights about the system. For instance, we observe that the ratio of the coefficient of training and the coefficient of age reduces with $n$, indicating that with every new stimulus application the impact of age on memory formation becomes stronger. 
The advantage of the prediction is its simplicity.
Moreover, the prediction is an ad-hoc description of the parameter dependence and other choices are possible; see Appendix E.
The stimulation protocol is characterised by the three parameters $n$, $t^\mathrm{train}$, and $t^\mathrm{wait}$, while the almost identical $\tau_\mathrm{pre}$ and $\tau_\mathrm{train}$  capture the characteristic time scale of network adaptation.

Finally, we investigate how many stimuli an adaptive network can store.
We demand that a stored stimulus can be read out at a later time, implying that its signal exceeds a given threshold $S_\mathrm{thresh}$, which captures uncertainties in the read-out apparatus as well as intrinsic noise.
Fig.~\ref{fig:4}(c) shows the fraction of stimuli that can be retrieved (where $S_n > S_\mathrm{thresh}$) as a function of the total number of stimuli, $N$, and the training time~$t^\mathrm{train}$.
In this case, large training times are detrimental since they age the network too much for later stimuli to be retrieved. Fig.~\ref{fig:4}(d) shows the number of stimuli that can be retrieved as a function of the training and waiting time. 
The largest number of stimuli is stored for smaller waiting time, as this reduces the age of the network.
 To find the optimal parameters for storing memory independent of a read-out apparatus specific threshold $S_{\mathrm{thresh}}$, we quantify the minimum signal out of the five stimuli's signals for varying training and waiting time, see Fig.~\ref{fig:4}(e). We observe that while the optimal waiting time is $0$, a non-zero optimal training time exists for storing memory.  
Taken together, our analysis reveals the strong trade-off between writing stimuli for a sufficient duration and the resulting inevitable ageing of the network that suppresses signals of subsequent stimuli.

\section{Discussion}
We showed that adaptive flow networks can store memory of multiple stimuli in the morphology of weak links, which cannot be revived in our model.
Consequently, signatures of earlier stimuli are not destroyed by subsequent evolution, in contrast to the behaviour of typical mechanical networks~\cite{Stern2019}.
Since older networks contain fewer strong links, which could shrink to store memory, the readout signal of each stimulus strongly decreases with the age of the network before the stimulus was written, which is similar to the memory plasticity observed in disordered system~\cite{Keim2011,Paulsen2014}.
Conversely, the signal strength increases with its training time, i.e., the duration the stimulus is presented, similar to memory formation by directed aging in mechanical networks~\cite{Hexner2020,Pashine2019}.
Taken together, we showed that adaptive flow networks reach maximal capacity at an intermediate training time, which compromises between imprinting sufficiently and ageing minimally.

Our work focuses on the simple situation that non-overlapping stimuli are subsequently applied at the edge of  flow networks of similar morphology.
To describe, realistic living flow networks, like \textit{Physarum} or our vasculature, our work will need to be extended in multiple directions:
First, the overall network geometry will have an impact on how stimuli are stored.
Work in mechanical networks~\cite{Rocks2019, Stern2019} suggests that the internal timescales of flow networks and their memory capacity will depend on network size.
Second, realistic systems deal with time-varying and potentially overlapping stimuli of various strengths.
Third, living systems can grow and expand \cite{Ronellenfitsch2016}, implying that links can possibly regrow from their minimal size and new links can be added to the network.
Taken together, it is likely that realistic adaptive flow networks show a dynamic behaviour, storing information about stimuli on various time scales.

Taken together, our work quantifies the memory capacity of adaptive flow networks for varying parameters, crucial for understanding the emergent behaviour of non-neuronal organism \textit{P.~polycephalum}.

\begin{acknowledgments}
This work was supported by the Max Planck Society. This project has received funding from the European Research Council (ERC) under the European Union’s Horizon 2020 research and innovation program (Grant Agreement No. 947630, FlowMem).
\end{acknowledgments}

\appendix
\section{Model of adaptive flow network} \label{sec:1}
In this section. we give a detailed derivation of the iteration rule for our model of adaptive networks.
A networks is modelled as a connected set of straight tubes described by a graph with $\nodes$ nodes enumerated by $i=1,2,\ldots,\nodes$ and tube $ij$ of length $l_{ij}$, which connect node $i$ and $j$.
Assuming Poiseuille flow in all tubes, conductances $C_{ij}$ are assigned to all tubes.
To study the flow in the network, we designate the first node ($i = 1$) as the outlet, while all other nodes ($i > 1$) are inlets with inflow $q_i(t)$.
Kirchhoff's current law then reads $L(t)\vec p(t) = \vec q(t)$, where $\vec p(t)=\{p_1(t), p_2(t), \ldots, p_{\nodes}(t)\}$ represents the pressures at all nodes, $\vec q(t)=\{q_1(t), q_2(t), \ldots, q_{\nodes}(t)\}$ represents the inflows at all nodes, and $L$ is the graph Laplacian matrix with elements $L_{ij} = \delta_{ij}\sum_{n}C_{in} - C_{ij}$, where $C_{ij}=0$ if tube $ij$ is not present.
We use Kirchoff's current law to find the instantaneous flow rate $Q_{ij}(t) = C_{ij}(t)(p_i(t) - p_j(t))$ through tube $ij$ at every time step $t$. 
Following \cite{Murray1926,Corson2010,Katifori2010,Bohn2007}, we evolve the network by adapting its conductances to minimise the power loss of the network, 

\begin{equation}\label{Power}
    E(t) = \sum_{<ij>} \frac{\mean{Q_{ij}(t)^{2}}_{T}}{C_{ij}(t)}
    \;,
\end{equation}
while obeying the constraint of the total network volume,
\begin{equation}\label{constraint}
    \mathcal{K}^{1/2} = \sum_{<ij>}(C_{ij}(t)l_{ij})^{1/2}l_{ij}
    \;.
\end{equation}
The optimal conductance $C_{ij}^{\star}(t)$ that minimise \Eqref{Power} at $t$, maintaining the constraint of available material given by \Eqref{constraint}, 
\begin{equation}\label{final}
   C_{ij}^{\star}(t) = \frac{\langle Q_{ij}(t)^2\rangle_T^{2/3}\mathcal{K}}
   {\left(\sum _{<ij>}\langle Q_{ij}(t)^{2}\rangle_T^{1/3}l_{ij}\right)^{2}l_{ij}}
	\;.
\end{equation}
The conductances $C_{ij}(t+\delta t)$ at time step $t+\delta t$ adapt to the optimal conductance $C_{ij}^{\star}(t)$
\begin{equation}
 C_{ij}(t+\delta t) \leftarrow C_{ij}^{\star}(t) \;.   
\end{equation}
We then repeat this rule iteratively over many time steps.
Taken together, this iterative rule for adapting conductances follows from independently optimising conductances $C_{ij}$ to minimises the power loss of the network  while keeping the total network volume fixed. 

In this work we model the network as an hexagonal network, with an addition of noise to the node positions to reduce artifacts generated from the regularities of the network. The node at the center of the network, $i=1$, is the sole outlet, with all the nodes in the network are sources of inflows which fluctuate by either turning \textit{on} $q_{i > 1} = 2q^{(0)}$ or \textit{off} $q_{i> 1} = 0$ with equal probability. Meaning the average base inflow through the source nodes are $<q_{i > 1}> = q^{(0)}$.
\section{Signal of last stimulus changes with its training and relaxation time}\label{sec:2}
\begin{figure*}[ht]
    \centering
    \includegraphics[width=\textwidth]{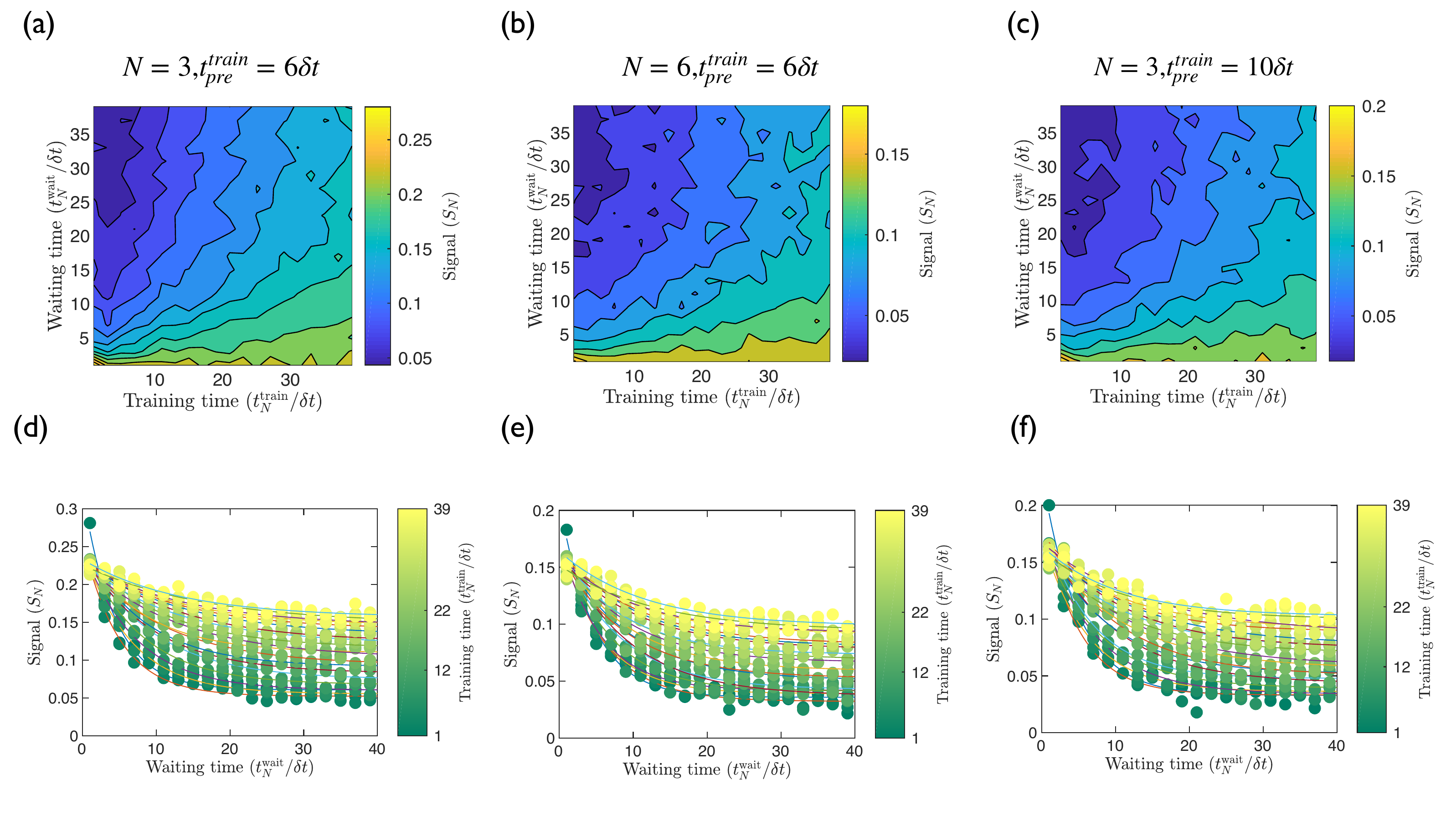}
    \caption{Signal $S_N$ of  final stimulus depend on its training time~$t^\mathrm{train}_N$ and waiting time~$t^\mathrm{wait}_N$.
    (a--c) $S_N$ for varying of $t^{\mathrm{train}}_N$ and $t^{\mathrm{wait}}_N$ for three different parameter sets.
    (d--f) The same data $S_N$ is plotted against $t^{\mathrm{wait}}_N$ for different $t^{\mathrm{train}}_N$ (colors).
    Lines represent exponential fits.
    (a--f) Model parameters: $t^{\mathrm{wait}}_{\mathrm{pre}} = 5\delta t$,$N_\mathrm{node}=760$,$q^\mathrm{add} = 2000q^{(0)}$, $q^{(0)} =1$, $T=30$, $\mathcal{K}=1600$.}
    \label{fig:s1}
\end{figure*}
We observe that the memory signal~$S_N$ of the last stimulus increases with its training time~$t^\mathrm{train}_N$ and decreases with its waiting time~$t^\mathrm{wait}_N$ independent of the pre-stimulation protocol, see Fig~\ref{fig:s1}.
This behaviour is similar to the signal of a stimulus written in a freshly initiated untrained network \cite{Bhattacharyya2022}. 
In particular, $S_N(t^{\mathrm{wait}}_N)$ can be fitted with exponential decay functions, which is similar to our observation in the main manuscript of $S_1$ decaying with the age after the stimulus application, $t_{\mathrm{post}}^{\mathrm{age}}$, following an exponential decay, while $S_1$ increases with the training time $t_{1}^{\mathrm{train}}$. Since, here the age after stimulus application is $t_{\mathrm{post}}^{\mathrm{age}} = t_N^{\mathrm{wait}}$. These observations imply a general dependency of signal $S_n$ on age, post stimulus application $t_{\mathrm{post}}^{\mathrm{age}}$, and training time of that particular stimulus, $t_{n}^{\mathrm{train}}$, for all $n$.
\section{Stable thick tubes do not decay irreversibly between stimuli}
\begin{figure*}[t]
    \centering
    \includegraphics[width=\textwidth]{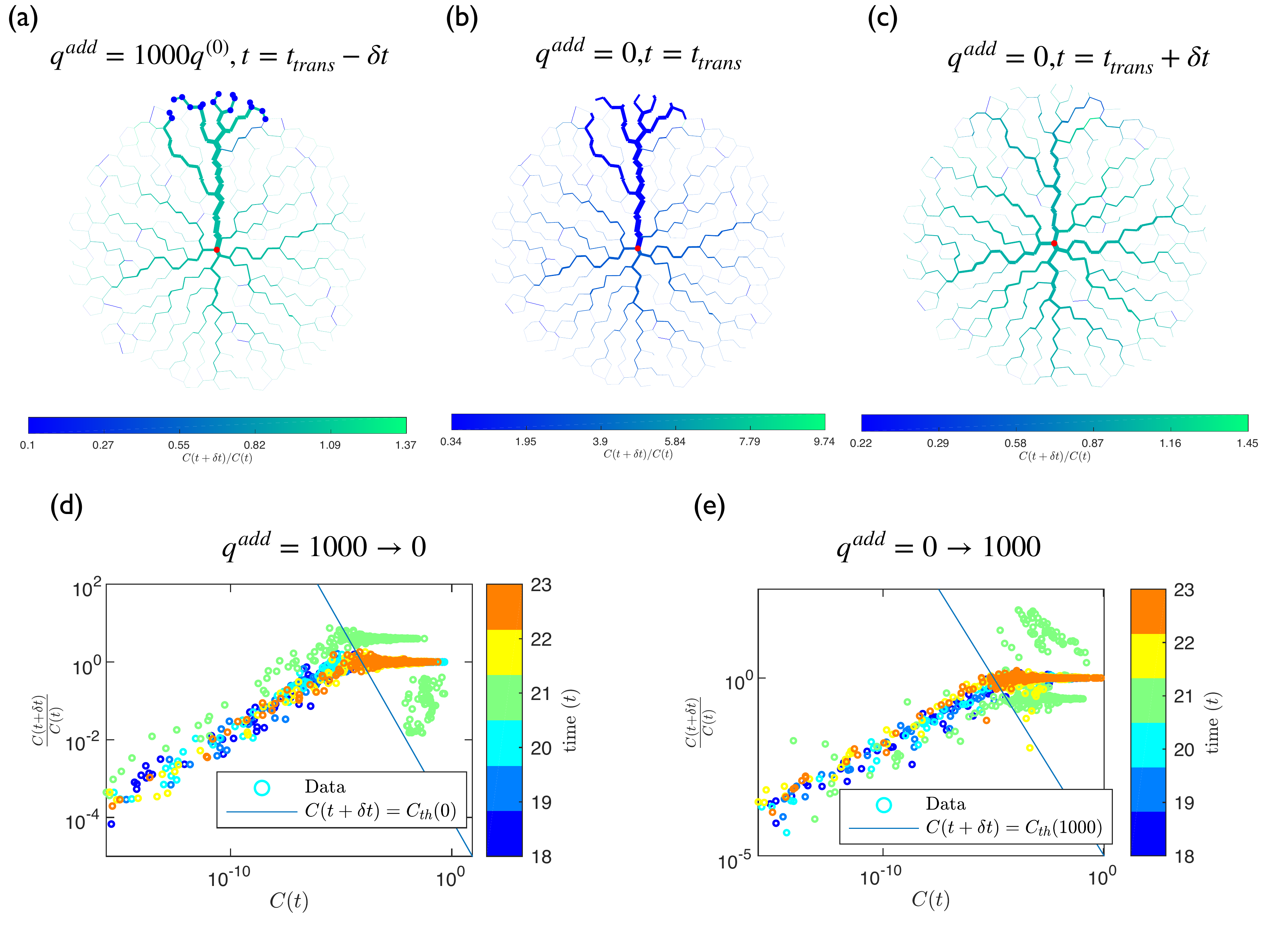}
    \caption{Dynamics of networks when a stimulus is enabled or disabled.
    (a--c) Network morphology at three consecutive time steps during which a stimulus is switched off. Color code indicates relative change of conductances, $C(t+\delta t)/C(t)$.
    (d, e) $C(t+\delta t)/C(t)$ as a function of $C(t)$ for $6$ different time point around the transition time $t=21$ where the stimulus is switch off (panel d) or switch on (panel e). 
    Model parameters are $\nodes =560$, $T=30\delta t$, and $q^{(0)} =1$.}
    \label{fig:s2}
\end{figure*}
To understand how adaptive networks evolve between stimuli, we here study the dynamics of conductances by measuring $C(t+\delta t)/C(t)$ for all the tubes of the network before and after the transition of switching the additional load from $q^\mathrm{add} = 1000q^{(0)}$ to $q^\mathrm{add} =0$.
We map the the dynamics of conductances on the network at $3$ different time step, $t=t_\mathrm{trans}-\delta t, t_\mathrm{trans}, t_\mathrm{trans}+\delta t$, where  $t_\mathrm{trans}$ is the transition time point when the stimulus is switched off.
Fig.~\ref{fig:s2}(a) shows that before the transition, the adaptive network shows a hierarchical network structure with thick tubes connecting the nodes where the stimulus is applied (blue nodes) to the outlet of the network (red node).
In particular, $C(t+\delta t)/C(t)$ of the thick tubes are approximately $1$, indicating that they do not evolve further.
After switching the stimulus off at the transition time at $t_\mathrm{trans}$, we notice that the conductances of the thick tubes start decaying, while other tubes grow; see Fig.~\ref{fig:s2}(b).
Already one step later, at $t_\mathrm{trans}+\delta t$, the network assumes an isotropic, tree-like structure; see Fig.~\ref{fig:s2}(c). 
The quantification in Fig.~\ref{fig:s2}(d, e) indicates that the conductances follow the universal dynamics of adaptive networks~\cite{Bhattacharyya2022}, where  conductances below a threshold conductance~$C_\mathrm{th}$ decay, while larger conductances fluctuate over time. 
The threshold value reads~\cite{Bhattacharyya2022}
\begin{equation}\label{CthFinal}
    C_\mathrm{th} = \frac{\mathcal{K}}{\left(1+\frac{\delta q}{q^{(0)}}\right)^{4}\left(\nodes+\frac{q^\mathrm{add}}{q^{(0)}}\right)^{4/3}}
    \;,
\end{equation}
where $\nodes$ denotes the number of nodes in a network, and $\delta q$ quantifies load fluctuations. 

The conductances show an unique dynamics at the exact time step when the stimulus is switched, at $t = t_\mathrm{trans} = 21$.
As an example, we observe a cluster of tubes with high conductances, either decaying see Fig.~\ref{fig:s2}(d) or growing see Fig.~\ref{fig:s2}(e). 
As we know from \cite{Bhattacharyya2022} that all the conductance below the threshold $C_\mathrm{th}$ given by \Eqref{CthFinal} will decay irreversibly, we plot the $C(t+\delta t) = C_\mathrm{th}$ line (the blue line in Fig.~\ref{fig:s2}(d) and (e)) to track the tubes that will decay irreversibly. Here, $C_{th}$ is obtained from fitting the dynamics of thin tubes to,
\begin{equation}
    \frac{C(t+\delta t)}{C(t)} = \Big[ \frac{1}{C_{th}}\Big]^{1/3}C(t)^{1/3},
\end{equation}
for a constant $q^{add}$ following \cite{Bhattacharyya2022}. In Fig.~\ref{fig:s2}(d), the line corresponds to $C_{th}(q^{add} = 0)/C(t)$ vs $C(t)$, and in Fig.~\ref{fig:s2}(e) the line corresponds to $C_{th}(q^{add} = 1000)/C(t)$ vs $C(t)$. Indicating all the data points with $C(t+\delta t)/C(t)$ higher than this line have $C(t+\delta t) > C_{th}$ after the transition.

We observe that the data points belonging to the cluster of high conductance tubes (at $t=21$) are above of the line, indicating although these tubes decay during the transition of switching the stimulus to $0$, yet does not decay irreversibly, see Fig.~\ref{fig:s2}(d).

These observations imply that, the tubes with very large conductances do not decay irreversibly due to the transitions of stimulus strength ($q^{add}$), and are retained in the network.
\section{Thin tubes cannot be recovered between stimuli}
We next attempt to understand the dynamics of smaller tubes during the transition between stimuli.

\begin{figure}[t]
    \centering
    \includegraphics[width=0.5\textwidth]{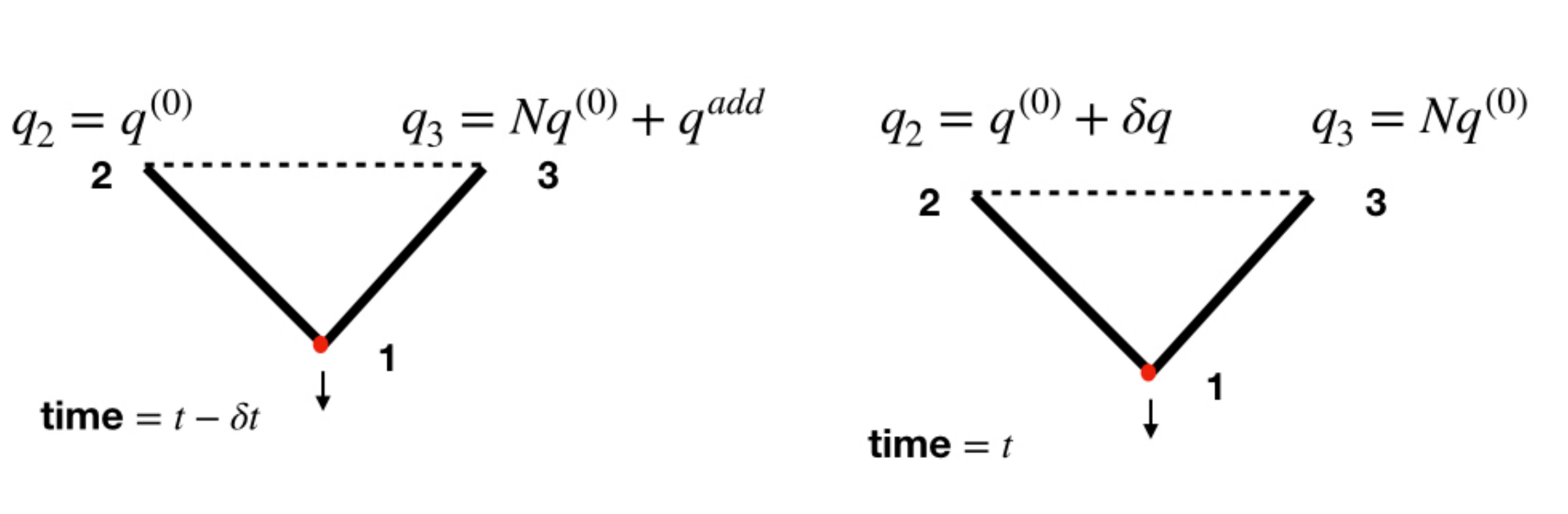}
    \caption{Minimal asymmetric network during the transition step, where the stimulus of addition is inflow switched off.
    (a) Minimal network at $t-\delta t$ where the inflow at node 2 is $q^{(0)}$ and at node 3 is $\nodes  q^{(0)}+q^\mathrm{add}$.
    (b) Minimal network at time $t$, where the inflow at node 2 is $q^{(0)}+\delta q$ and node 3 is $\nodes q^{(0)}$.}
    \label{fig:s2a}
\end{figure}

For this, we calculate the tubes' dynamics in the minimal triangular network shown in Fig.~\ref{fig:s2a}(a), which follows the same adaptation rule as the simulated adaptive networks,
\begin{align}\label{main23}
    C_{ij}(t+\delta t) &= Q_{ij}(t)^{4/3}.\frac{\mathcal{K}}{A(t)^2}
& \text{with} &&    
    A(t) &= 
    \sum_{<i,j>} Q_{ij}(t)^{2/3}
    \;.
\end{align}
We focus on the dynamics when $q^\mathrm{add} = 1000 \rightarrow 0$, but we expect the opposite case of $q^\mathrm{add} = 0 \rightarrow 1000$ to exhibit similar  behaviour. 
For the minimal network, we can express Kirchhoff's current law, $L \vec{p} = \vec{q} $, explicitely
\begin{equation}
    \left[ {\begin{array}{c}
        q_{1}  \\
        q_{2} \\
        q_{3} \\
   \end{array}}\right]=
  \left[ {\begin{array}{ccc}
   C_{13}+C_{12} & -C_{12} & -C_{13} \\
   -C_{12} & C_{12}+C_{23} & -C_{23} \\
   -C_{13} & -C_{23} & C_{13}+C_{23} \\
  \end{array} } \right]
  \left[ {\begin{array}{c}
        p_{1}  \\
        p_{2} \\
        p_{3} \\
   \end{array}}\right]
   \;;
\end{equation}
see Section~\ref{sec:1}.
Setting $p_{1} = 0$ because node 1 is a sink, we have
\begin{equation}
\left[ {\begin{array}{cc}
    C_{12}+C_{23} & -C_{23} \\
   -C_{23} & C_{13}+C_{23} \\
  \end{array} } \right]^{-1}
    \left[ {\begin{array}{c}
        q_{2} \\
        q_{3} \\
   \end{array}}\right]=
  \left[ {\begin{array}{c}
        p_{2} \\
        p_{3} \\
   \end{array}}\right]
\end{equation}
and thus
\begin{multline}
   \frac{1}{C_{13}(C_{12}+C_{23})+C_{12}C_{23}}
\left[ {\begin{array}{cc}
    C_{13}+C_{23} & C_{23} \\
   C_{23} & C_{12}+C_{23} \\
  \end{array} } \right]\\
    \left[ {\begin{array}{c}
        q_{2} \\
        q_{3} \\
   \end{array}}\right] =
  \left[ {\begin{array}{c}
        p_{2} \\
        p_{3} \\
   \end{array}}\right]\;. 
\end{multline}

Solving for $p_2$ and $p_3$, we find
\begin{equation}\label{P2n3}
\begin{split}
        p_{2} &= \frac{1}{C_{13}(C_{12}+C_{23})+C_{12}C_{23}}\Big( (C_{13}+C_{23})q_{2} + C_{23}q_{3}\Big)\\
        p_{3} &= \frac{1}{C_{13}(C_{12}+C_{23})+C_{12}C_{23}}\Big( C_{23}q_{2} + (C_{12}+C_{23})q_{3}\Big)
        \;.
\end{split}
\end{equation}
We assume the network has been adapted with a stimulus of an additional inflow and reached a local energy minimum at $t-\delta t$, implying a tree morphology  which connects all inlets directly to the outlet.
We can then assume the inflow on node $2$ and $3$ is distributed through tubes ${12}$ and ${13}$, respectively. 
Without loss of generality, we assume the additional load has been applied to node 3, implying $Q_{21}(t-\delta t) = q_{2}(t-\delta t) = q_{2}$ and $Q_{31}(t-\delta t) = q_{3}(t-\delta t) = q_{3}+q^\mathrm{add}$. 
Using \eqref{main23}, we can then approximate the conductances of tube $12$ and $13$ as $C_{12}(t) = q_{2}(t-\delta t)^{4/3}\mathcal{K}/A(t-\delta t)^2$ and $C_{13}(t) = q_{3}(t-\delta t)^{4/3}\mathcal{K}/A(t-\delta t)^2$, respectively, where we assumed $C_{23}(t)/C_{12}(t) \ll 1$ and $C_{23}(t)/C_{13}(t) \ll 1$ since $Q_{23}(t-\delta t) \ll Q_{21}(t-\delta t)$ and $Q_{23}(t-\delta t) \ll Q_{31}(t-\delta t)$.

During the transition from $t-\delta t$ to $t$, the stimulus is switched from a non-zero value to $0$, implying the load at node $3$ reduces to $q_{3}(t)  = q_{3}$.
The load at node $2$ also changes to $q_{2}(t) = q_{2}+\delta q$ due to the background fluctuation of magnitude  $\delta q$, where we assume this perturbation to be small compared to the load at node 2, $\delta q/q_{2} \ll 1$. 
In general, the minimal network is connected with a bigger network with $\nodes$ nodes, 
so the calculation we present in the following subsections considers the simple case of an asymmetric connection, with highest inflow difference,
as shown in the Fig.~\ref{fig:s2a}(a),
\begin{equation}\label{q23asym}
    q_{2} = q^{(0)},\: q_{3} \approx \nodes  q^{(0)} 
    \;,
\end{equation}
which implies tube $12$ is thinner than $13$, $C_{12}(t)< C_{13}(t)$, in the steady state network.
\subsection{Dynamics of medium thick tubes}
The flow rate $Q_{12}$ through tube ${12}$ reads
\begin{multline}\label{currQ12}
        Q_{12} (t) = C_{12}(p_{2}-p_{1})\\ = C_{12}(t)\frac{[C_{13}(t)+C_{23}(t)]q_{2}(t) + C_{23}(t)q_{3}(t)}{C_{13}(t)(C_{12}(t)+C_{23}(t))+C_{12}(t)C_{23}(t)}
\end{multline}
where we used \Eqref{P2n3}.
Inserting the values of $C_{13}(t)$, $C_{12}(t)$, $q_{2}(t)$, and $q_{3}(t)$ obtained above, we find 
\begin{equation}\label{Q12}
    Q_{21}(t) =  C_{12}(t)\frac{(q_{3}+q^\mathrm{add})^{4/3}(q_{2}+\delta q)}{\Big(q_{2}(q_{3}+q^\mathrm{add})\Big)^{4/3}}\cdot\frac{A(t-\delta t)^2}{\mathcal{K}}
    \;.
\end{equation}
Using \Eqref{Q12} in \Eqref{main23}, we can write
\begin{multline}
        C_{12}(t+\delta t) 
        = C_{12}(t)^{4/3}\frac{(q_{3}+q^\mathrm{add})^{16/9}(q_{2}+\delta q)^{4/3}}{\Big(q_{2}(q_{3}+q^\mathrm{add})\Big)^{16/9}} \\ \cdot \frac{A(t-\delta t)^{8/3}}{\mathcal{K}^{4/3}}\frac{\mathcal{K}}{A(t)^2}
\end{multline}
and
\begin{equation}
        \label{bigveincalc}
        \frac{C_{12}(t+\delta t)}{C_{12}(t)} = \left(1+\frac{\delta q}{q_{2}}\right)^{4/3}\frac{\Big(q_{2}^{2/3}+(q_3+q^\mathrm{add})^{2/3}\Big)^{2}}{\Big((q_{2}+\delta q)^{2/3}+q_{3}^{2/3}\Big)^{2}}
        \;.
\end{equation}
Expanding in the small parameter $q^{(0)}/q^\mathrm{add}$ results in 

\begin{equation}\label{MediumThick}
    \frac{C_{12}(t+\delta t)}{C_{12}(t)} = \frac{(\nodes +\frac{q^\mathrm{add}}{q^{(0)}})^{4/3}}{\nodes ^{4/3}} = \left(1 + \frac{q^\mathrm{add}}{\nodes q^{(0)}}\right)^{4/3}
    \;,
\end{equation}
which implies the medium thick tubes $12$ will grow when the stimulus is switched off.
\subsection{Dynamics of thin tubes with largest inflow difference}
We next calculate the dynamics of the thin tube $23$ of the minimal network shown in Fig.~\ref{fig:s2a}.
The flow rate~$Q_{23}$ through tube ${23}$ follows from \Eqref{P2n3}, 
\begin{equation}\label{currQ23}
        Q_{23} (t)= C_{23}(p_{2}-p_{3}) = C_{23}\frac{C_{13}q_{2} - C_{12}q_{3}}{C_{13}(C_{12}+C_{23})+C_{12}C_{23}}
        \;.
\end{equation} 
Inserting the values of $C_{13}(t)$, $C_{12}(t)$, $q_{2}(t)$, $q_{3}(t)$, we find
\begin{multline}\label{currentt+1}
    Q_{23}(t) = C_{23}(t)\frac{(q_{3}+q^\mathrm{add})^{4/3}(q_{2}+\delta q) - (q_{2})^{4/3}(q_{3})}{\Big( q_{2}(q_{3}+q^\mathrm{add})\Big) ^{4/3}} \\
    \cdot \frac{A(t-\delta t)^2}{\mathcal{K}}
    \;.
\end{multline}
Using \Eqref{main23},  we write the conductance $C_{23}(t+\delta t)$ at time $t+\delta t$ as a function of conductance $C(t)$ at time $t$,
\begin{equation}\label{middlestep}
    \frac{C_{23}(t+\delta t)}{C_{23}(t)} = \mathcal{S}_\mathrm{disable}   C_{23}(t)^{1/3}
    \;,
\end{equation}
where the prefactor 
reads
\begin{multline}\label{alpha1}
\mathcal{S}_\mathrm{disable} = \frac{\Big( (q_{3}+q^\mathrm{add})^{4/3}(q_{2}+\delta q) - (q_{2})^{4/3}(q_{3}) \Big)^{4/3}}{\Big( q_{2}(q_{3}+q^\mathrm{add})\Big) ^{16/9}}\\ \cdot \frac{A(t-\delta t)^{8/3}\mathcal{K}}{\mathcal{K}^{4/3}A(t)^2}   
\;.
\end{multline}
Since $A(t-\delta t) = \Big(q_{2}^{2/3}+(q_{3}+q^\mathrm{add})^{2/3}\Big)$ and $A(t) = \Big( (q_{2}+\delta q)^{2/3}+q_{3}^{2/3} \Big)$,
the prefactor becomes

\begin{multline}\label{alpha2}
\mathcal{S}_\mathrm{disable} = \frac{\Big( (q_{3}+q^\mathrm{add})^{4/3}(q_{2}+\delta q) - (q_{2})^{4/3}(q_{3}) \Big)^{4/3}}{\Big( q_{2}(q_{3}+q^\mathrm{add})\Big) ^{16/9}}\\ \cdot \frac{\Big( q_2^{2/3} +(q_3+q^\mathrm{add})^{2/3}\Big)^{8/3}}{\mathcal{K}^{1/3}\Big((q_2+\delta q)^{2/3}+q_3^{2/3}\Big)^2}   
\;.
\end{multline}
Inserting the values of $q_2$ and $q_3$ from\Eqref{q23asym}, we obtain

\begin{multline}
    \mathcal{S}_\mathrm{disable} =
    	\frac{\Big( \Big[\frac{1}{\nodes+\frac{q^\mathrm{add}}{q^{(0)}}}\Bigr]^{2/3}+1\Big) ^{8/3}}{\mathcal{K}^{1/3}} \\
	\cdot \frac{\Big( \frac{1}{\nodes}(\nodes+\frac{q^\mathrm{add}}{q^{(0)}})^{4/3}(1+\frac{\delta q}{q^{(0)}})-1\Big) ^{4/3}}{\Big( (\frac{1}{\nodes}(1+\frac{\delta q}{q^{(0)}}))^{2/3}+1\Big) ^{2}}
	\;.
\end{multline}
Assuming a large network, $\nodes q^{(0)} + q^\mathrm{add} \gg q^{(0)}$, we find
\begin{multline}\label{SFinal}
    \mathcal{S}_\mathrm{disable}
    \approx \frac{(1+\frac{\delta q}{q^{(0)}})^{4/3}\Big( \nodes + \frac{q^\mathrm{add}}{q^{(0)}}\Big) ^{16/9}}{\nodes^{4/3}\mathcal{K}^{1/3}}\\
    = \frac{(1+\frac{\delta q}{q^{(0)}})^{4/3}\Big(\nodes+\frac{q^\mathrm{add}}{q^{(0)}}\Big)^{4/9}\Big(1+\frac{q^\mathrm{add}}{\nodes q^{(0)}}\Big)^{4/3}}{\mathcal{K}^{1/3}}
    \;.
\end{multline}
Since tubes with $C(t+\delta t) < C_\mathrm{th}(q^\mathrm{add}=0)$ will decay irreversibly after the stimulus is disabled, we find
\begin{equation}
    \mathcal{S}_\mathrm{disable} C(t)^{4/3 }< C_\mathrm{th}(q^\mathrm{add}=0)
\end{equation}
and thus
\begin{equation}
    C(t) < \left(\frac{C_\mathrm{th}(q^\mathrm{add}=0)}{S_\mathrm{disable}}\right)^{3/4}
    \;.
\end{equation}
Using $C_\mathrm{th}$ from \Eqref{CthFinal} and $\mathcal{S}_\mathrm{disable}$ from \Eqref{SFinal}, we find
\begin{multline}
    \left(\frac{C_\mathrm{th}(q^\mathrm{add}=0)}{S_\mathrm{disable}}\right)^{3/4}
    	= \Big( \frac{\mathcal{K}}{(1+\frac{\delta q}{q^{(0)}})^{4}\nodes^{4/3}} \\ \cdot \frac{\mathcal{K}^{1/3}}{(1+\frac{\delta q}{q^{(0)}})^{4/3}\Big(\nodes+\frac{q^\mathrm{add}}{q^{(0)}}\Big)^{4/9}\Big(1+\frac{q^\mathrm{add}}{\nodes q^{(0)}}\Big)^{4/3}}\Big)^{3/4},\\
     = \frac{K}{(1+\frac{\delta q}{q^{(0)}})^{4}(\nodes+\frac{q^\mathrm{add}}{q^{(0)}})^{4/3}} = C_\mathrm{th}(q^\mathrm{add})
    \;.
\end{multline}

This implies tubes with $C(t) > C_\mathrm{th}(q^\mathrm{add})$ will be retained in the network even after disabling the stimulus.
Moreover, tubes that were decaying in the presence of the stimulus will continue decaying even when the stimulus is disabled.
\begin{figure*}
    \centering
    \includegraphics[width=\textwidth]{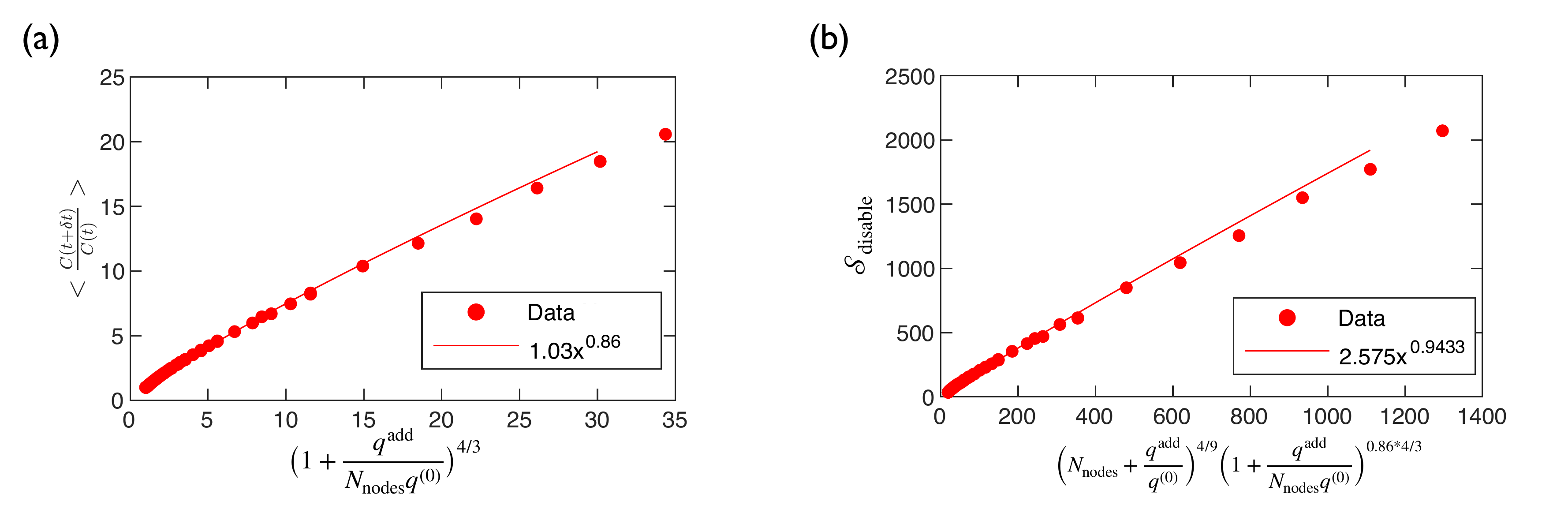}
    \caption{Dynamics of tube conductances during the transition of disabling a stimulus matches analytical approximations.
    (a) Average conductance changes $\mean{C(t+\delta t)/C(t)}$ of medium thick tubes in networks as a function of the corresponding analytical prediction given by \Eqref{MediumThick}.
    The solid line is a power-law fit.
     (b) The pre-factor $\mathcal{S}_\mathrm{disable}$, obtained from fitting the dynamics of thin tubes to $C(t+\delta t)/C(t) = \mathcal{S}C(t)^{1/3}$, as a function of the prediction given by \Eqref{SFinal}.
     , including the correction of the power obtained from (a), shows the data follows a straight line with the corrected prediction.
     (a--b) For varying $q^{(0)}$ and $q^{\mathrm{add}}$ and fixed model parameters $\nodes = 760$, $T=30\delta t$. }
    \label{fig:s3}
\end{figure*}
Fig.~\ref{fig:s3} shows that our analytical prediction is reasonably correct.
However, we observe that the measured value of the dynamics of the medium thick tubes are slightly lower than that of the analytical prediction, probably because the calculation only considers the most extreme case of highest inflow difference. 
Taken together, we show the established hierarchy of tube conductances cannot be erased by switching the stimulus from a non-zero value to $0$ or vice versa. 

\section{Memory signal from combination of age and training components}
\begin{figure*}[ht!]
    \centering
    \includegraphics[width = \textwidth]{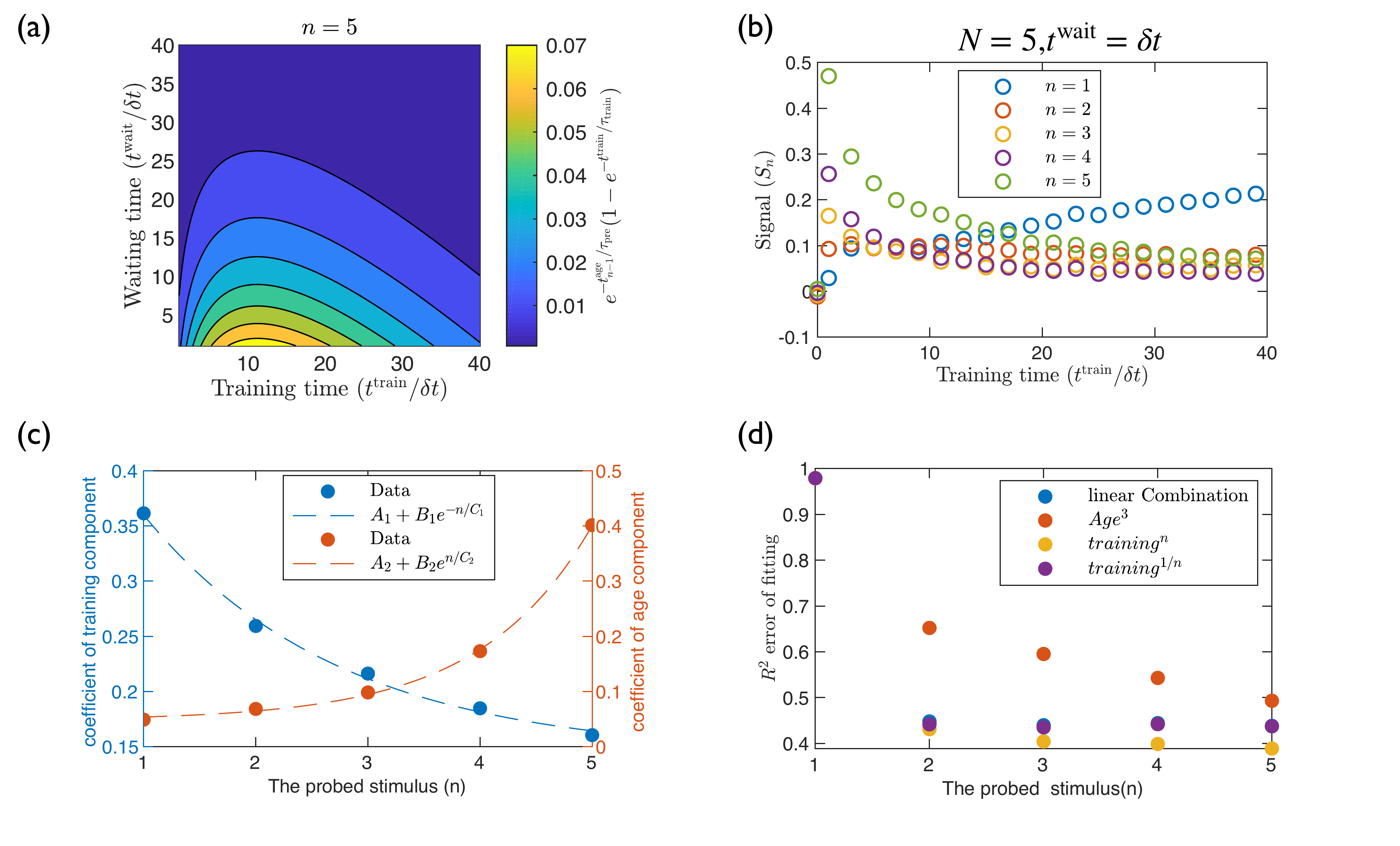}
    \caption{Memory signal~$S_n$ depends on training time $t^\mathrm{train}$ and age of the network before training.
    (a)~Analytical prediction assuming $S_n$ is a product of the impact of age and training.
    (b)~$S_n$ as a function of $t^\mathrm{train}$ for various~$n$.
    (c)~Fit parameters (coefficient of training and age component) obtained while fitting memory readout signal as a sum of training and age component, plotted against the probed stimulus. (d) $R^2$ of the fit obtained from fitting the signal as various combination of training and age with respect to the probed stimulus.}
    \label{fig:S5}
\end{figure*}
In the main text, we observed the signal~$S_1$ of the first stimulus increases with the training time $t^{\mathrm{train}}$,
\begin{equation}
   S_1(t^{\mathrm{train}}) \sim (1-e^{-t^{\mathrm{train}}/\tau_{\mathrm{train}}})
   \;. 
\end{equation}
We also observed that the signal~$S_N$ of the last stimulus decays with the age $t^{\mathrm{age}}_{N-1}$ before the stimulus application,
\begin{equation}
    S_N(t^{\mathrm{age}}_{N-1}) \sim e^{-t^{\mathrm{age}}_{N-1}/\tau_{\mathrm{pre}}}
    \;.
\end{equation}
Combining these observations, we hypothesise that the signals~$S_n$ of general stimuli between the first and last stimulus are affected by both effects.
$S_n$ should thus be a function of $S^{\mathrm{train}}_n=(1-e^{-t^{\mathrm{train}}_n/\tau_{\mathrm{train}}})$ and $S^{\mathrm{age}}_n=e^{-t^{\mathrm{age}}_{n-1}/\tau_{\mathrm{pre}}}$.

We start by asking whether a simple product is an adequate description,
\begin{equation}\label{AnalMul}
    S_n = S^{\mathrm{age}}_n  S^{\mathrm{train}}_n = e^{-t^{\mathrm{age}}_{n-1}/\tau_{\mathrm{pre}}}(1-e^{-t^\mathrm{train}_n/\tau_{\mathrm{train}}})
    \;.
\end{equation}
Fig.~\ref{fig:S5}(a) and Fig.~\ref{fig:S5}(b) shows that this prediction cannot reproduce the qualitative dependency of the signal on $t^{\mathrm{train}}$ and $t^{\mathrm{wait}}$.
In particular, \Eqref{AnalMul} suggests a maximal signal for a non-zero training time for any $n$, while the numerical data shown in Fig.~4 of the main text shows that the signal is strongest for very small $t^{\mathrm{train}}$ and $t^{\mathrm{wait}}$,  specifically for $n>3$.
\begin{figure*}[p]
    \centering
    \includegraphics[width = \textwidth]{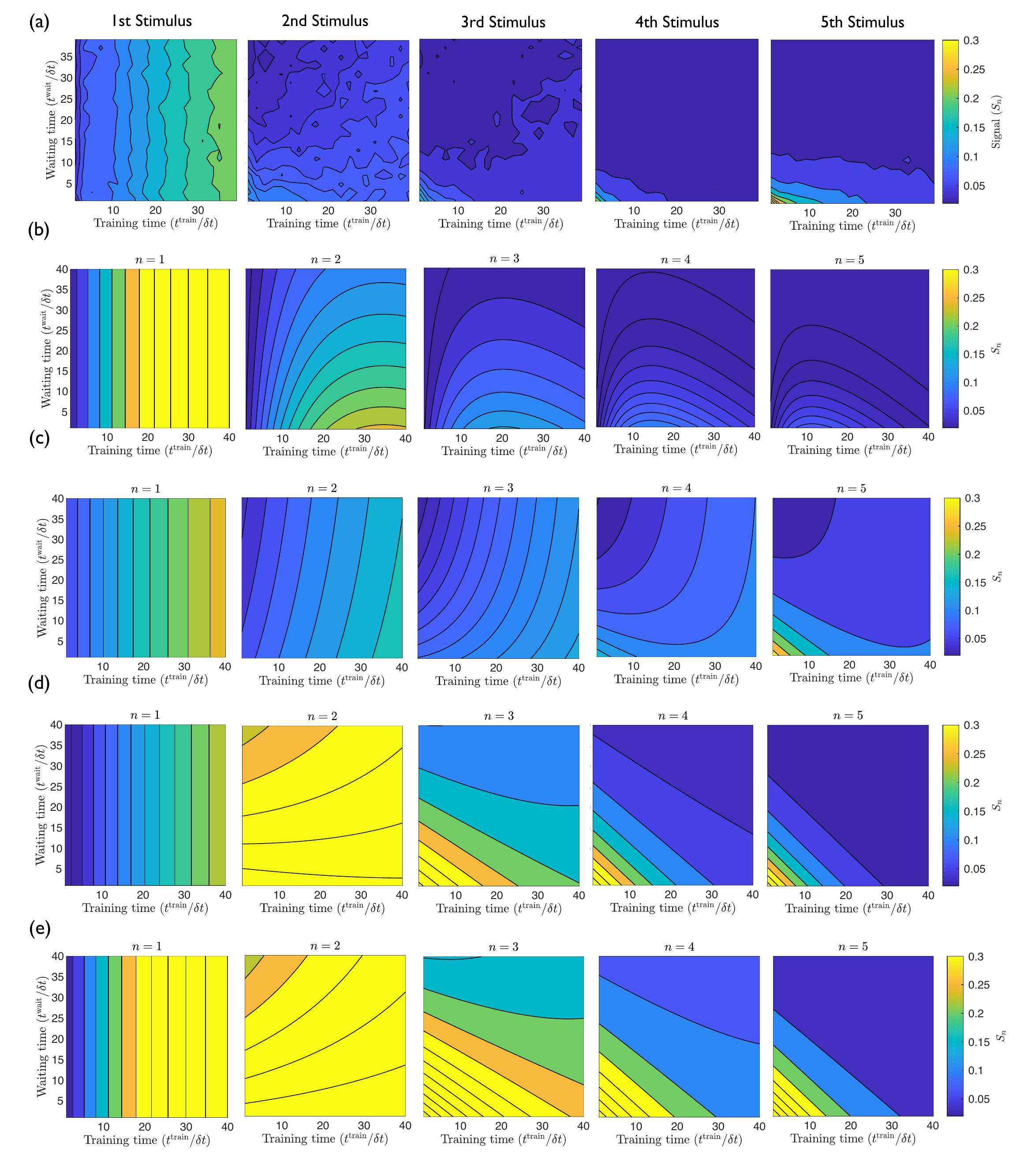}
    \caption{Analytical approximation of signal w.r.t training and waiting time when nth stimulus is probed
    (a) Numerically obtained memory read out signal with respect to training and waiting time as shown in Fig.~4(a)
    (b) Signal $S_n$ approximated as a product of $S_n^{\mathrm{age}}$ and $S_n^{\mathrm{train}}$, in \Eqref{AnalMul}, with parameters $\tau_{\mathrm{pre}}=50\delta t$ and $\tau_{\mathrm{train}}=50\delta t$.
    (c) Signal $S_n$ approximated as a linear sum of $S_n^{\mathrm{age}}$ and $S_n^{\mathrm{train}}$ as shown in \Eqref{Approx1} using the functional form found from fitting the data \Eqref{Num}, with parameters $\tau_{\mathrm{pre}}=50\delta t$ and $\tau_{\mathrm{train}}=50\delta t$. 
    (d) $S_n = \frac{1}{2}\big( e^{-t^{\mathrm{age}}_{n-1}/\tau_{\mathrm{pre}}} + \frac{n}{2}^{-n+1}(1-e^{-t^{\mathrm{train}}/\tau_{\mathrm{train}}})\big)$, with parameter $\tau_{\mathrm{pre}} =50\delta t$ and $\tau_{\mathrm{train}} = 62\delta t$ (e) $S_n = (1-e^{-(n+1)/1.5})e^{-t^{\mathrm{age}}_{n-1}/\tau_{\mathrm{pre}}} +e^{-(n-1)/1.5}(1-e^{-t^{\mathrm{train}}/\tau_{\mathrm{train}}})$ with parameters $\tau_{\mathrm{pre}}=50\delta t$ and $\tau_{\mathrm{train}}=50\delta t$.}
    \label{fig:S6}
\end{figure*}
 We next hypothesis that $S_n$ is a linear combination of $S^{\mathrm{age}}_n$ and $S^{\mathrm{train}}_n$,
\begin{equation}\label{Approx1}
    S_n = f^{\mathrm{train}}_n S^{\mathrm{age}}_n + f^{\mathrm{age}}_n S^{\mathrm{train}}_n
    \;,
\end{equation}
where we allow different coefficients $f^{\mathrm{train}}_n$ and $f^{\mathrm{age}}_n$ for each stimulus.
We determine these by fitting $S_n/S^{\mathrm{age}}_n(t^{\mathrm{train}},t^{\mathrm{wait}})$ to $f^{\mathrm{age}}_n + f^{\mathrm{train}}_n S^{\mathrm{train}}_n/S^{\mathrm{age}}_n(t^{\mathrm{train}},t^{\mathrm{wait}})$ for different values of $n$ over a range of $t^{\mathrm{train}}/\delta t \in [1, 40]$ and $t^{\mathrm{wait}}/\delta t \in [1,40]$.
Fig.~\ref{fig:S5}(c) shows that a suitable fit is given by
\begin{equation}\label{Num}
    \begin{split}
    f^{\mathrm{train}}_n &= 0.14+0.386 e^{-n/1.75}\\
        f^{\mathrm{age}}_n &= 0.0475+0.0023e^{n/0.99}
    \end{split}
\end{equation}
We note in Fig.~\ref{fig:S5}(d) that the $R^2$ error of the fit to compute the coefficients of each component is $\approx 0.4$, indicating that the signal is not just a linear superposition of the components. We use a few different models to fit the data. As example, we fit the signal to $S_n= f_n^{\mathrm{train}}S_n^{\mathrm{train}} + f_n^{\mathrm{age}}(S_n^{\mathrm{age}})^{3}$ as we do not expect the power of the age component to change with n. Additionally we fit the signal to some other simple non-linear composition of age and training component, as example $S_n = f_n^{\mathrm{train}}(S_n^\textrm{train})^{n} + f_n^{\mathrm{age}}S_n^\textrm{age}$ and $S_n = f_n^{\mathrm{train}}(S_n^\textrm{train})^{1/n} + f_n^{\mathrm{age}}S_n^\textrm{age}$. However, none of these models increase the $R^2$ of the fit significantly. So for simplicity, we choose to approximate the signal as a linear sum of $S^{\mathrm{train}}_n$ and $S^{\mathrm{age}}_n$. We observe, if we choose the coefficients of $S^{\mathrm{train}}_n$ and $S^{\mathrm{age}}_n$ following \eqref{Num}, the signal of a stimulus obtained using this approximation qualitatively agrees with the numerical observation, see Fig.~\ref{fig:S6}(c).

Now to obtain a generic analytical approximation of the memory read-out signal without including many different parameters, that also describes the signal as a linear superposition of the age and training impact, we assume that the coefficient of $S^{\mathrm{age}}_n$ does not change and the coefficient of $S^{\mathrm{train}}_n$ decreases of n. 
We assume these coefficients because, \Eqref{Num} suggests that the coefficient of the training component reduces with $n$ and the coefficient of age component increase with $n$. Moreover, the fit parameters show that the change of the coefficient of age component over n ($\sim 0.0023$) is much smaller than the change of the coefficient of the training component over n ($\sim 0.36$), which agrees with our observation in Fig.~2(b).

Using these observations we now choose the following bounded function which is a sum of $S_n^{\mathrm{age}}$ and $S_n^{\mathrm{train}}$ to approximate the signal of $n$th stimulus when $n>1$.
\begin{equation} \label{Approxa}
\begin{split}
    S_{n} &= e^{-t^{\mathrm{age}}_{n-1}/\tau_{\mathrm{pre}}} + f^{\mathrm{train}}_n(1-e^{-t^{\mathrm{train}}/\tau_{\mathrm{train}}}),\\
    &= \frac{1}{2}\big( e^{-t^{\mathrm{age}}_{n-1}/\tau_{\mathrm{pre}}} + (\frac{n}{2})^{-n+1}(1-e^{-t^{\mathrm{train}}/\tau_{\mathrm{train}}})\big).
\end{split}
\end{equation}
Additionally, from the numerical observation, we approximate for $n=1$, 
\begin{equation}
    S_1 = \frac{1}{2}(1-e^{-t^{\mathrm{train}}/\tau_{\mathrm{train}}}).
\end{equation}
Although \Eqref{Approxa} reproduces, the dependency of signal on training and waiting time qualitatively (see, Fig.~4(b) and Fig.~\ref{fig:S6}(d)), this does not reproduce the dependency quantitatively. 
However, we show that the same qualitative dependency on training and waiting time is robust to small changes of parameters. As example, when \Eqref{Approxa} is plotted assuming $\tau_{\mathrm{pre}} = 50\delta t$ and $\tau_{\mathrm{train}} =62\delta t$, see Fig.~\ref{fig:S6}(d), the qualitative feature does not change from the observation in Fig.~4(b). Similarly the qualitative dependency does not change, when different functions $f^{\mathrm{train}}_n$ and $f^{\mathrm{age}}_n$ are chosen as the coefficients of the training and age impact. As an example, if the chosen functions are,
\begin{equation}
\begin{split}
        f^{\mathrm{train}}_n = e^{-(n-1)/1.5},\\
        f^{\mathrm{age}}_n = (1-e^{-(n-1)/1.5}),
\end{split}
\end{equation}
consequently, the signal of the nth stimulus is approximated as,
\begin{multline}\label{Approx2}
    S_n = (1-e^{-(n-1)/1.5})e^{-t^{\mathrm{age}}_{n-1}/\tau_{\mathrm{pre}}}\\ +e^{-(n-1)/1.5}(1-e^{-t^{\mathrm{train}}/\tau_{\mathrm{train}}}),
\end{multline}
the qualitative dependency of approximated signal on $t^{\mathrm{train}}$ and $t^{\mathrm{wait}}$ is similar to Fig.~4(b), see Fig.~\ref{fig:S6}(e).
Note that in both approximations \Eqref{Approxa} and \Eqref{Approx2}, the ratio of the coefficients of age and training impact increases with n, meaning after a certain number of pre-stimuli the impact of age dominates over the impact of training.


%

\end{document}